\documentstyle[preprint,aps,floats]{revtex}
\input psfig.tex
\begin{document}
\draft
%\twocolumn[\hsize\textwidth\columnwidth\hsize\csname @twocolumnfalse\endcsname
\preprint{PUPT-95-1569, astro-ph/9510072}
\title{Looking for $\Lambda$ with the Rees-Sciama Effect}
\author{Robert G.  Crittenden and Neil Turok}
\address{Joseph Henry Laboratory, Princeton University,\\
Princeton NJ, 08544.}
\date{10/12/95}
\maketitle

\begin{abstract}
In models with a cosmological constant,
a significant component of the large scale 
cosmic microwave background 
(CMB) anisotropy 
is produced 
at rather low redshifts, 
$z \mathrel{\raise.3ex\hbox{$<$}\mkern-14mu\lower0.6ex\hbox{$\sim$}} 1$.
In these models, 
the gravitational potential perturbations begin to evolve
at late times. 
Photons 
passing through these time varying potentials aquire 
frequency shifts, an effect first noted by Rees and Sciama.
Since the gravitational potential is associated 
with matter at observable redshifts,
the local matter density (or some tracer
thereof) is correlated with the CMB anisotropy on the sky. We
outline the optimal method of measuring 
this correlation, and discuss the prospects for
using an X-ray/COBE comparison to detect a cosmological constant. 

\end{abstract}
\hspace{0.2in}
%]
\section{Introduction}

The idea of a cosmological constant ($\Lambda$) has been a recurring one 
ever since Einstein first proposed it \cite{einstein20}.
Recent motivations for a non-zero $\Lambda$ include 
easing the `age crisis,'  reconciling dynamical measures of the matter 
density with prejudices for flatness and 
increasing the power in large scale perturbations \cite{recent}.
But rather than simply introducing another free parameter, 
it is more interesting to ask whether 
there are specific observational
signals that could confirm or refute the hypothesis of
a nonzero $\Lambda$. 

We propose one such test here,
which uses the fact that a $\Lambda$ term causes 
the Newtonian potential $\Phi$ to start evolving at late times, 
producing a significant amout of CMB anisotropy \cite{ks85}.
Since $\Lambda$ comes to dominate rather suddenly, this effect is
most important at rather modest redshifts. 
But if 
observations of the density field allow us 
to reconstruct the local potential, then this 
should 
be correlated with the microwave sky. 
Measuring this correlation thus would constrain $\Lambda$.

The strongest present observational constraint on $\Lambda$ is that from
gravitational lensing, which results from the fact that if there 
were a large cosmological constant, 
then lensing events would be seen more frequently
than they are.
The handful of lensing events 
that have been observed constrain 
the fraction of the critical density contributed by $\Lambda$ to be, 
$\Omega_\Lambda < 0.7$ \cite{maozrix93}. 
This constraint, however, is sensitive to how well the mass distributions 
of early 
type galaxies are modeled and relies on the assumption that no lensing events 
are obscured by dust.  
Other probes of $\Lambda$, such as measurements of the deceleration 
parameter $q_0$, give weaker constraints \cite{cpt92}.
Whether our test becomes competitive with these remains to be seen, 
but the types of biases in the various tests are so different that 
it is worth exploring them all.

\section{R-S effect}

In the approximation of
instantanous recombination,
the microwave anisotropy in a direction ${\bf n}$ on the sky is
given by the formula
\begin{equation}
\frac{\delta T}
{T}({\bf{n}})
 = \bigl[{1\over 4} \delta_\gamma   + {\bf v\cdot n} +\Phi \bigr]_i^f
+2 \int_i^f d\tau \dot{\Phi} (\tau, {\bf n}(\tau_0-\tau)).
\label{eq:esw}
\end{equation}
The integral is over
the conformal time $\tau$, $\tau_f=\tau_0$ being today and $\tau_i$ being
recombination.
The first term represents the perturbations on the
surface of last scattering, namely the perturbation to the density of
the radiation-baryon fluid ($\delta_\gamma$), the Doppler term 
(${\bf v\cdot n}$),
 and the Newtonian potential.
The second term, usually called the Integrated Sachs-Wolfe (ISW) term, 
represents the effect of a time varying gravitional potential
along the line of sight. Heuristically, it represents the red shifting 
of photons which must `climb out' of  a different potential than they
`fell into'.  This is called the
Rees-Sciama effect \cite{rs68}.

In a flat, matter dominated universe, with linear growing 
density perturbations, $\Phi$ is constant and there is no
Rees-Sciama effect.
Nonlinear gravitational collapse does lead to anisotropies on very small 
angular scales, but of small amplitude 
\cite{Seljak}. 
In a universe with a signifcant cosmological constant, however, 
$\Phi$ becomes time dependent even in linear theory
and an appreciable amount of anisotropy can be created at quite modest 
redshifts.

As $\Lambda$ increases, it comes to dominate the energy density at 
earlier and earlier redshifts.  
The effect on the evolution of the potential is thus more pronounced, 
as is the corresponding anisotropy generated at late times.
For smaller values of $\Lambda$ the opposite is true; the correlated 
anisotropy is less, but it is more concentrated at very late epochs.
As an aside, we should note that 
the $\Lambda$ also has an indirect effect on the 
degree scale anisotropy, because in a flat universe 
the presence of $\Lambda$ 
alters the matter-radiation balance at last scattering.
In contrast, the large scale  Rees-Sciama
effect is independent of physics at high redshifts (e.g. reionization). 

To quantify this,
we expand the sky temperature in the usual spherical 
harmonics
\begin{equation}
\frac{\delta T}
{T}({\bf{n}})
\equiv \sum_{l,m}a_{lm} Y_{lm}(\theta, \phi),
\end{equation}
where in an isotropic ensemble the $a_{lm}$ obey 
$\langle a_{lm} a_{l'm'}\rangle = \delta_{l l'} \delta_{m m'} C_l$ with 
$C_l$ the angular power spectrum.
An idea of how much anisotropy is produced from the late time
evolution of $\Phi$ is obtained by computing the contribution 
to each $C_l$ 
by the ISW integral prior to some redshift $z_c$. This is 
shown in Figure 1.
From this we see that a significant fraction of the 
$C_l$'s at low $l$ are produced at 
$z \mathrel{\raise.3ex\hbox{$<$}\mkern-14mu\lower0.6ex\hbox{$\sim$}} 1$.

\begin{figure}[htbp]
\centerline{\psfig{file=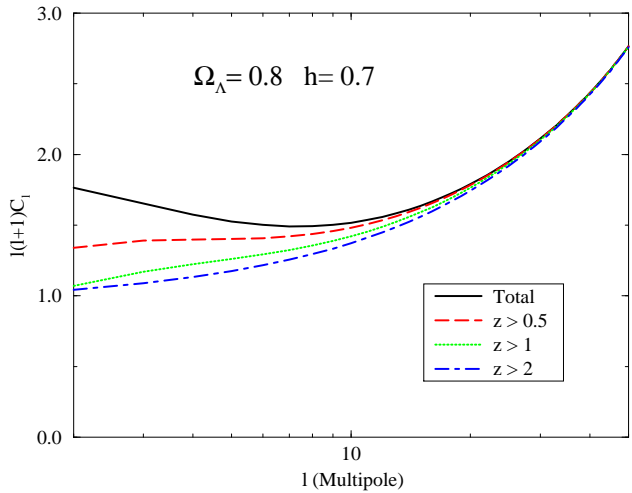,width=3.9in}}
\caption{
The large scale anisotropy power spectrum,  $C_l = \langle |a_{lm}|^2
\rangle$, for  a model with $\Omega_\Lambda = 0.8$ and $h = 0.7$.  
Also shown is the anisotropy that is produced prior to a given 
red shift, for $z = 0.5, 1,$ and $2$.  
A significant portion of the anisotropy is produced rather recently. 
}
\label{fig:f1}
\end{figure}

\section{Correlation with density}

Since part of the CMB anisotropy is associated with the 
gravitational potential at low redshift, it must be correlated 
with the matter distribution in our vicinity. The gravitational potential
is determined from the matter distribution by
Poisson's equation $\nabla^2 \Phi 
= 4\pi G a^2 \delta_M \rho_M $, where $\delta_M $ is the fractional density
perturbation in the matter and $\rho_M$ is the background matter density. 

It is convenient to treat this in Fourier space, 
so that for example  $\Phi({\bf x}, \tau)
 =\sum_{\bf k} \Phi({\bf k}, \tau) e^{i{\bf k\cdot x}}$, and also to 
refer the density perturbation to the present time $\tau_0$. 
In the matter dominated epoch, all $k$ modes grow at the
same rate, and from Poisson's equation one infers that
$\Phi({\bf k}, \tau) = g(\tau) \delta_M({\bf k}, \tau_0)/k^2$,
where $g(\tau)$ is independent of $k$. 
Inserting this in relation
(\ref{eq:esw}), and expanding the plane wave in 
spherical Bessel functions one finds
\begin{equation}
a_{lm}^{RS} = 8 \pi i^l \sum_{\bf k} Y_{lm}^*(\Omega_{\bf k})
 {\delta_M({\bf k}, \tau_0) \over k^2}\int d \tau \dot{g}(\tau) 
j_l(k \Delta \tau),
\label{eq:rs}
\end{equation}
where $\Delta \tau=\tau_0-\tau$. This equation has a simple interpretation
in real space: it says that the R-S contribution to 
$a_{lm}$ comes from convolving the matter density $\delta_M({\bf x}, \tau_0)$
perturbation in our vicinity with a spatial weighting function 
$f_{l}(r)Y_{lm}^*(\Omega_{\bf x})$.
That is, if we substitute the inverse Fourier transform, we find
\begin{equation}
a_{lm}^{RS} = \int d^3 {\bf x} f_{l}(r) \delta_M({\bf x},\tau_0)
Y_{lm}^*(\Omega_{\bf x}),
\label{eq:fa}
\end{equation}
where
\begin{equation}
 f_l(r) = \int d\tau \dot{g}(\tau) \int { dk \over \pi^2} 
j_l(kr) j_l(k\Delta \tau).
\label{eq:fb}
\end{equation}

The integral is straightforwardly performed, with the result that
\begin{equation}
f_l(r) = {2^{2l} \over \pi (2l+1)} 
\int d\tau \dot{g}(\tau)  {(r\Delta \tau)^l \over 
(r+\Delta \tau +|r-\Delta \tau|)^{2l+1}} .
\label{eq:fc}
\end{equation}
Equations (\ref{eq:fb}) and (\ref{eq:fc}) tell one how to compute the
Rees-Sciama contribution to each $a_{lm}$. 
The asymptotics of $f$ are easily read off: as $r\rightarrow 0$,
$f \rightarrow const$, and as $r\rightarrow \infty$, $f \sim
r^{-(l+1)}$. More importantly, $f_l(r)$ is reasonably described 
by a very simple approximation:
for large $l$ (we shall only be interested in $l>2$) the second term
in the integral is approximately a delta function $\delta 
(\Delta \tau -r)$,
and the integral is approximately
\begin{equation}
 f_l(r) \simeq  {1 \over 2\pi l(l+1)}\dot{g}(\tau_0 -r),
\end{equation}
i.e., it is proportional to the rate of 
change of the local gravitational potential. 
We have checked that this is a reasonable approximation
down to $l=2$.
Figure 2 shows $\dot{g}(z) = \dot{g}(\tau_0 - \tau(z))$ 
for a range of values for $\Lambda$.
Note that
$f_l(r)$ is independent of the power spectrum of primordial 
density perturbations. The only assumption needed is that 
the perturbations are in the pure growing mode.

\begin{figure}[htbp]
\centerline{\psfig{file=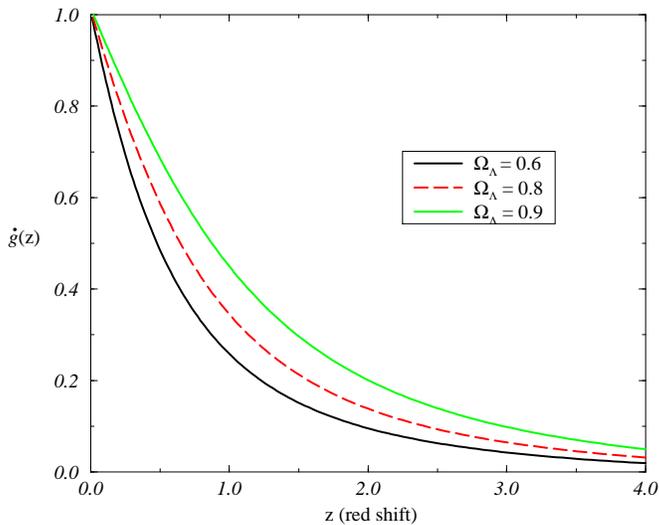,width=3.9in}}
\caption{
The ideal filter function, $\dot{g}(z)$, as a function of red shift. 
Even for very large $\Lambda$, significant contributions 
result from low redshift, though contributions begin at higher redshifts. 
}
\label{fig:f2}
\end{figure}

The observed $a_{lm}$'s will differ from the Rees-Sciama 
result, however, because  
a significant component of the observed anisotropy is 
produced on the last scattering surface. 
The latter acts to obscure the correlation between 
the observed anisotropy and the local density fluctuations.
What sort of signal-to-noise may we ultimately expect in the 
cross correlation of the density and anisotropy,
given that we are limited by cosmic variance? 

\begin{figure}[htbp]
\centerline{\psfig{file=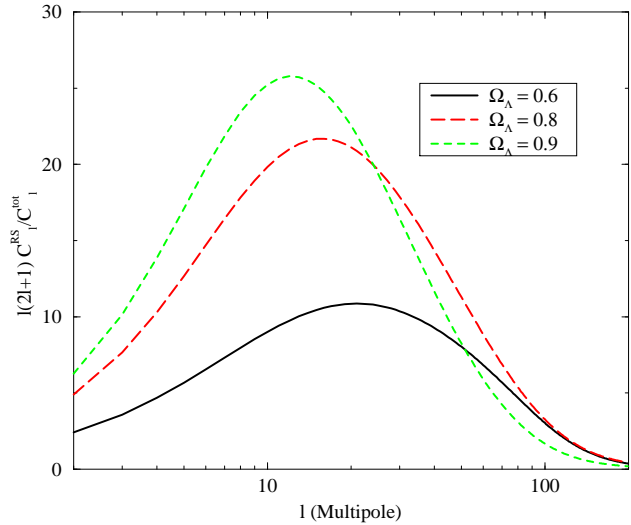,width=3.9in}}
\caption{
We plot the signal to noise squared as a 
function of $l$, where the area under the curve represents the contribution 
for a given logarithmic interval. 
}
\label{fig:f3}
\end{figure}

Let us begin with the most optimistic assumption, 
that we have a complete survey of some tracer of the matter distribution, 
deep enough to see all redshifts 
where the cosmological constant was significant.
We wish to compare the hypothesis that the $a_{lm}^{RS}$'s defined by
eq. (\ref{eq:fa})  {\it are} correlated with 
the observed $a_{lm}$'s as predicted by the  $\Lambda$ models,
with the hypothesis that they are not correlated at all.
The relative likelihood of the two hypotheses can be computed for
any given data set; if the correlations are real, then the expected value of
this is
\begin{equation}
{\cal P} =  \prod_{l,m} \left(1- 
{\langle a_{lm}^{RS} a_{lm}^{tot *}\rangle^2  \over
 C_l^{RS} C_l^{tot}}\right)^{-1}.
\label{eq:ppp}
\end{equation}
(For a set of independent observables, 
${\cal P}$ is the product of the individual ${\cal P}$'s). 
Defining the signal to noise ratio as ln(${\cal P}$)
we infer that 
\begin{equation}
\left({S\over N}\right)^2 \equiv  \ln {\cal P} 
 \geq 
\sum_{l} (2l+1){\langle a_{lm}^{RS} a_{lm}^{tot *}\rangle^2 
 \over  C_l^{RS} C_l^{tot}}.
\label{eq:fsn}
\end{equation}
This sum converges quickly beyond $l\sim 50$,
yielding $S/N \geq 5.5, 7.4,$ and $7.9$ for $\Omega_\Lambda= .6, 
.8,$ and $.9$ respectively.
Figure \ref{fig:f3} shows the contribution to this sum as a function of $l$. 
Note that the Rees-Sciama contribution is almost uncorrelated with
the remainder of the anisotropy, so that $\langle a_{lm}^{RS} a_{lm}^{tot *}
\rangle \approx  C_l^{RS}$.

\begin{figure}[htbp]
\centerline{\psfig{file=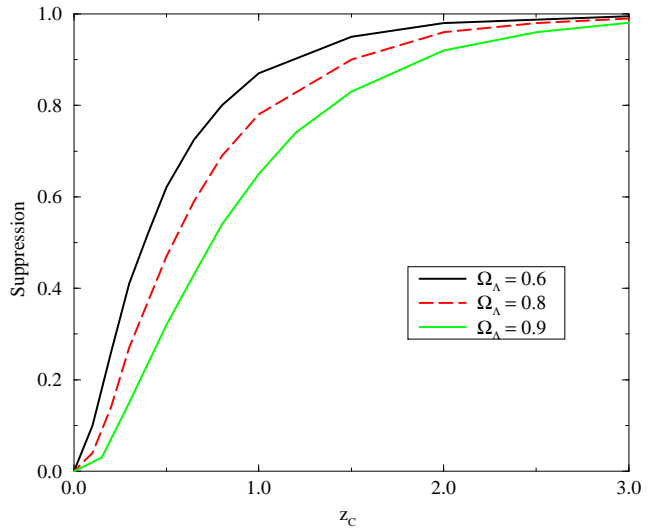,width=3.9in}}
\caption{
We plot the reduction of the signal to noise ratio if the 
density survey is cut off beyond a given redshift.
}
\label{fig:f4}
\end{figure}

Realistic surveys, however, are not likely to probe the density this deeply.
For a survey which is less than ideal, we can get some feel for 
the loss in signal by considering the case where 
the convolution function,  $f_l(r)$, is the ideal one 
out to some cutoff red shift, $z_c$,  and zero beyond.
The signal to noise ratio in the correlation for a given multipole is
then 
\begin{equation}
{S\over N}^{part}  \geq 
{\langle a_{lm}^{part} a_{lm}^{tot *}\rangle \over
\sqrt{ C_l^{part} C_l^{tot} }}  \approx
{\langle a_{lm}^{part} a_{lm}^{RS *}\rangle \over
\sqrt{ C_l^{part} C_l^{RS} }} {S\over N}^{ideal}.
\label{eq:partial}
\end{equation}
The suppression factor is given by
\begin{equation}
{\langle a_{lm}^{part} a_{lm}^{RS *}\rangle \over
\sqrt{ C_l^{part} C_l^{RS} }} = {\int k^2 dk f_l(k) \tilde{f}_l(k) P_k 
\over \sqrt{\int k^2 dk f_l(k)^2 P_k \int k^2 dk \tilde{f}_l(k)^2 P_k }}
\label{eq:partials}
\end{equation}
where,  
\begin{equation}
f_l(k) = \int r^2 dr j_l(kr) \dot{g}(r) \qquad
\tilde{f}_l(k) = \int_0^{z_c} r^2 dr j_l(kr) \dot{g}(r) 
\end{equation}
and $P_k \equiv \langle |\delta_M(k,\tau_0)|^2 \rangle$.
We have performed these integrals numerically and find that the result is 
very weakly dependent on $l$.
The resulting suppression factor for $l=10$ 
as a function of the redshift 
is shown in Figure \ref{fig:f4}. As can be seen, there 
is a substantial signal even when the survey is cut off at 
rather modest $z_c$. 

\section{Results}

To predict the $a_{lm}^{RS}$, 
one requires a measure of the density contrast $\delta_M$
in our vicinity. Traditionally, it is assumed that 
this is at least roughly proportional to the fluctuation 
in the number density $n(\bf{x})$ of galaxies (or other tracers): 
 $(\delta n/\bar{n}) = b (\delta \rho/\bar{\rho})$ where $b$ is a
`bias' factor which could depend on redshift.
The dimensionless cross correlation between the R-S anisotropy
level predicted from a survey of mass tracers
and the detected CMB anisotropy, i.e. 
$\langle a_{lm}^{pred} a_{lm}^{det}\rangle 
/ \sqrt{C_l^{pred}
C_l^{det}}$, is independent of $b$ if $b$ is constant,
but it does depend on 
the variation of $b$ with redshift. 
However, the net effect is to 
alter the effective window function $f_l(z)$,
and we have seen that 
the cross correlation is fairly insensitive to this. 
A real data analysis could 
set limits on the variation of $b(z)$ and on $\Lambda$, and
might be used to determine the magnitude of $b$,  
should a correlation be found.

It would be very interesting to cross correlate
the ROSAT all-sky X-ray survey with the COBE anisotropy measurement. 
The X-rays with energies of order 
a keV appear to be consistent with a simple model in
which they are all produced by active galactic nuclei (Seyfert 
galaxies and quasars). Surveys of five deep fields to find these
AGNs indicate that their distribution in redshift (i.e. $d\bar{N}/dz$)
is approximately 
flat for $.5<z<2$ and cuts off rapidly thereafter \cite{comastri}, 
so they do indeed sample 
the redshift range of interest.
To estimate the expected correlation, however, we need to translate this
into an effective weighting function $f_l(r)$.

At any frequency, the intensity of the X-ray sky in a given direction is 
$\iota({\bf n})
 = \int {\cal F}(z) dN(r{\bf n},z)$ ,
where ${\cal F}(z)$ is the mean flux from a source at redshift $z$ and 
$dN(r{\bf n},z)$ is the number of sources in the redshift 
interval $[z,z+dz]$.  (Here $r = \tau_0-\tau(z)$.)  We can express $dN$ as, 
$dN(r{\bf n},z) = d \bar{N}(1+b(z)\delta_M(r{\bf n},z))$, where 
$d\bar{N}$ is the mean value of $dN$.
We  then obtain
\begin{equation}
\delta \iota({\bf n})
 = \int dz {d\bar{N}\over dz} {\cal F}(z) b(z) D(z) \delta_M(r{\bf n},z=0),
\end{equation}
where $D(z)$ is the matter growth factor
normalized to unity today. 
Comparing this with equation (4), 
we can identify 
\begin{equation}
r^2 f_l(r) \propto 
b(z)  {dz\over d\tau} {d\bar{N}\over dz} D(z) {\cal F}(z),
\end{equation}
thus giving us an expression for the actual experimental weighting function.
Using a simple fit to the $d\bar{N}/dz$ given in \cite{comastri},
and the naive assumptions 
that $b(z)$ and ${\cal F}(z) $ are constant, 
we find a suppression factor of $\sim 0.8$  for an 
$\Omega_\Lambda =0.8$ universe. 
Barring other sources of noise, a substantial signal should be 
visible 
in the COBE-ROSAT correlation, at least for this model. 

Very recently the large angular scale fluctuations in the
ROSAT survey have been studied, with the finding that there 
is a significant autocorrelation on scales $\theta <6^o$ \cite{soltan}. 
(It is argued that
this could be accounted for if 
$\sim 30 $ \% of the X-ray background fluctuations were due to a new class 
of X-ray sources in Abell clusters). 
Unfortunately, the effect we are discussing would be 
most visible on scales larger than this; 
but even if
the autocorrelation of the ROSAT survey is statistically insignificant
on these scales, it is possible that the cross-correlation with
COBE is still significant (just as was the 
case for the FIRS experiment \cite{FIRS}). 

In closing, we wish to emphasize that a correlation between 
the local density perturbations and the CMB anisotropy should exist 
in most cosmological models to some degree. The time independence 
of the Newtonian potential in the flat matter dominated universe 
is very much a special case. In particular we expect a similar 
effect in spatially open universes and in models with cosmic defects.

We thank Jim Peebles for suggesting the use of ROSAT for this test, 
and Ue-Li Pen and Ed Turner for helpful discussions.
This work was partially supported by NSF contract
PHY90-21984, and the David and Lucile Packard Foundation.

\end{document}